\documentclass[twocolumn,aps,prl,showpacs,floats,letterpaper,floatfix,groupedaddress]{revtex4}

\usepackage{amssymb,amsmath}
\usepackage[dvips]{graphicx}

\usepackage{dcolumn,epsfig}

\newcommand{\beq}{\begin{equation}}
\newcommand{\eeq}{\end{equation}}
\newcommand{\bes}{\begin{subequations}}
\newcommand{\ees}{\end{subequations}}
\newcommand{\bea}{\begin{eqnarray}}
\newcommand{\eea}{\end{eqnarray}}
\newcommand{\ba}{\begin{array}}
\newcommand{\ea}{\end{array}}
\newcommand{\beqn}{\begin{eqnarray*}}
\newcommand{\eeqn}{\end{eqnarray*}}

\newcommand{\f}[2]{\frac{#1}{#2}}

\newcommand{\la}{\langle}
\newcommand{\dg}{\dagger}
\newcommand{\ra}{\rangle}

\newcommand{\bk}{\bf {k}}

\def\nn{\nonumber}

\newlength{\sizeonefig}
\newlength{\sizetwofig}
\begin{document}

\title{Two-photon scattering by a driven three-level emitter in a one-dimensional waveguide and electromagnetically induced transparency}

\author{Dibyendu Roy} 
\affiliation{Department of Physics, University of California-San Diego, La Jolla, California 92093-0319, USA}

\begin{abstract}
We study correlated two-photon transport in a (quasi) one-dimensional photonic waveguide coupled to a three-level $\Lambda$-type emitter driven by a classical light field. Two-photon correlation is much stronger in the waveguide for a driven three-level emitter (3LE) than a two-level emitter. The driven 3LE waveguide shows electromagnetically induced transparency (EIT), and we investigate the scaling of EIT for one and two photons. We show that the two transmitted photons are bunched together at any distance separation when energy of the incident photons meets ``two-photon resonance'' criterion for EIT. 
\end{abstract}

\vspace{0.5cm}

\maketitle

Realization of controlled optical nonlinearity at the level of individual photons would have direct application to build large scale quantum networks for optical quantum information processing \cite{Kimble08}. Unlike electrons, photons do not interact with each other, but photon-photon interaction can be created using strong light-matter interaction in optically thick media such as ultracold atomic gases. Although for practical implementation in quantum information processing one would need to produce optical nonlinearity with a few atoms and photons. With this inspiration, experimentalists have just recently demonstrated electromagnetically induced transparency (EIT) for a weak probe light field with a single natural or artificial atom in a cavity \cite{Mucke10, Kampschulte10, Abdumalikov10} or in free space \cite{Slodicka10}. EIT is a destructive quantum interference phenomenon where the presence of a control light field eliminates absorption of the probe light by a multilevel atom \cite{Harris97, Harris90, Fleischhauer05}. Thus, the atom acts as an optical transistor where transmission of a light is controlled by another light field.    

Recently a new scheme to achieve strong photon-photon interaction in one-dimensional (1D) waveguides has been proposed \cite{Kojima03, Shena07, Shenb07,Chang07}. It has been shown in a system consisting of a two-level emitter (2LE) coupled to a 1D continuum for photons. One interesting feature of the 1D systems is that the spontaneously emitted and scattered waves from the 2LE will always interfere with the incident wave. Local interaction generates a strong correlation between photons in the waveguide by preventing multiple occupancy of photons at the 2LE. 
 Several nanoscale systems such as photonic crystal waveguides \cite{Faraon07}, surface plasmon modes of metallic nanowires \cite{Chang06, Akimov07}, microwave transmission lines \cite{Astafiev10}, optical nanofibers \cite{Dayan08, Vetsch10}, semiconductor or diamond nanowires \cite{Claudon10, Babinec10} would act as 1D continuum for photons. Now, it will be  interesting to study photon transport in a 1D waveguide with  a multi-level atom, which will show more complex quantum interference phenomena.  Single photon scattering by various configurations of a three-level emitter (3LE) in a quasi 1D waveguide has been evaluated in Ref.\cite{Witthaut10}. 
 The purpose of this Letter is to study correlated two-photon dynamics in a 1D waveguide coupled to a driven $\Lambda$-type 3LE.   

\begin{figure}
\includegraphics[width=6.0cm]{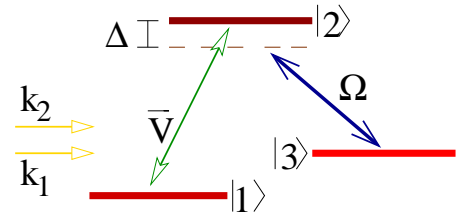}
\caption{Schematic of a three-level emitter driven by a classical laser beam with Rabi frequency $\Omega$, and two incident photons with wave number $k_1$, $k_2$ from left.}
\label{pl1}
\end{figure}
We consider the Hamiltonian for the full quasi 1D system as $ \mathcal{H}=\mathcal{H}_{0}+\mathcal{H}_{3LE}+\mathcal{H}_c$, where $\mathcal{H}_{0}$ and $\mathcal{H}_{3LE}$ represent, respectively, the free photons and the 3LE in the waveguide. $\mathcal{H}_c$ denotes local coupling between the photon modes and the 3LE. We consider a linear energy ($E_{k}$)--momentum ($k$) dispersion for the free photons, i.e., $E_{k}=v_gk$, where $v_g$ is the group velocity of the photons. We divide the positive and the negative momentum photons as right- and left-moving modes. Thus we write $ \mathcal{H}_{0}=-iv_g \int dx [a^{\dg}_R(x)\partial_x a_R(x)- a^{\dg}_L(x)\partial_x a_L(x)],$ where $a_R(x)~ [a_L(x)]$ is the annihilation operator of a right-[left-] moving photon at position $x$. We set $v_g=1$ for simplicity. We consider that the excited state $|2\ra$ of the $\Lambda$-type emitter (Fig. \ref{pl1}) 
is coupled to another level $|3\ra$ by a classical laser beam with Rabi frequency $\Omega$ and detuning $\Delta$. The energy of the ground state $|1\ra$ is set to zero as reference. The 3LE Hamiltonian within the rotating wave approximation is $\mathcal{H}_{3LE}=(E_2-i\gamma_2/2)|2\ra\la2|+(E_2-\Delta-i\gamma_3/2)|3\ra\la3|+(\Omega/2)(|3\ra\la2|+|2\ra\la3|),$
where spontaneous emission from states $|2\ra$ and $|3\ra$ to other modes out of the 1D waveguide is modeled by including an imaginary part $-i\gamma_2/2$ and $-i\gamma_3/2$ to the energies of the excited levels in the 3LE \cite{Witthaut10, Carmichael93}. Here we take the 3LE being side coupled to the propagating light modes locally at $x=0$; then $\mathcal{H}_c=\bar{V}|2\ra\la1|(a_R(0)+a_L(0))+{\rm H.c.},$
where $\bar{V}$ is the coupling strength between the 3LE and the photons. With the standard transformation to even-odd field operators, i.e., $a_{e}(x)=(a_{R}(x)+a_{L}(-x))/\sqrt{2}$ and $a_{o}(x)=(a_{R}(x)-a_{L}(-x))/\sqrt{2}$, the Hamiltonian of the full system breaks into two decoupled parts; i.e., $\mathcal{H}=\mathcal{H}_e+\mathcal{H}_o$ where 
$\mathcal{H}_e =-i\int dx~ a^\dg_{e}(x)\partial_xa_{e}(x) + \mathcal{H}_{3LE} + V \big(a^\dg_{e}(0)|1\ra\la2|+|2\ra\la1|a_{e}(0)\big)~{\rm and}~\mathcal{H}_o =-i\int dx~ a^\dg_{o}(x)\partial_ xa_{o}(x)$ with $V=\sqrt{2}\bar{V}$. 

Single-photon dynamics in this system has been studied recently in Ref.\cite{Witthaut10}. The single-photon transmission and reflection amplitudes are given by $\tilde{t}_{k}=\chi/(\chi+i\Gamma/2)$ and $\tilde{r}_k=-i\Gamma/2(\chi+i\Gamma/2)$ respectively, where $\Gamma=V^2$ and $\chi=E_k-E_2+i\gamma_2/2-\Omega^2/4(E_k-E_2+\Delta+i\gamma_3/2).$ The transmission coefficient $T_k$ and reflection coefficient $R_k$ show EIT-type line-shapes. $T_k$ becomes unity at the incident energy $E_k=E_2-\Delta$ [``two-photon resonance'' (TPR)] for the state $|3\ra$ being meta-stable, i.e., $\gamma_3=0$. The population of the excited state $|2\ra$ goes to zero at the TPR, this leads the 3LE to the ``dark state.''  The width of the transparency window depends only on the strength of the control field $\Omega$. With increasing value of $\gamma_3$ from zero, $T_k$ at the TPR falls from unity, and finally the EIT line shape is completely washed away.

The system of a 2LE coupled to a 1D continuum for photons is equivalent to a bosonic version of the celebrated single-impurity Anderson model for an infinite repulsive interaction between photons at the 2LE \cite{Shena07, Shenb07, Roy10}.  Two-photon transport in this system has been investigated recently for both side-coupled \cite{Shena07,Shenb07, Chang07} and direct-coupled emitter \cite{Roy10}, and it shows strong photon-photon interaction in higher order correlations of the transmitted and reflected fields. Here we carry out the two-photon transport for the driven $\Lambda$-type 3LE coupled to a 1D continuum of photons following Ref.\cite{Roy10} based upon the Bethe ansatz. It is expected that the two-photon dynamics  of the present model is much more complicated than for the 2LE due to complex interferences at the 3LE. We here solve the problem for a strong coupling $(\Gamma)$ in the presence of a classical laser beam of frequency $\Omega$ with $\Omega \ll\Gamma$. 
The two-photon incoming state with two incident photons from the left (i.e., right-moving) is given by
\bea
\int dx_1 dx_2 \f{1}{2\pi \sqrt{2}}\phi_{\bf k}(x_1,x_2)\f{1}{\sqrt{2}}a^{\dg}_R(x_1)a^{\dg}_R(x_2)|0,-\ra~, \label{instate}
\eea
where $\phi_{\bf k}(x_1,x_2)=(e^{ik_1x_1+ik_2x_2}+e^{ik_1x_2+ik_2x_1})$ with ${\bf k}=(k_1,k_2)$, and the total energy  of the two photons $E_{\bk}=k_1+k_2$. Applying the even-odd transformation for the field operators, we determine the different components of the two-photon incoming state into $ee,~oo$, and $eo$ subspaces. The general two-photon scattering eigenstate in the various subspaces is given by
\begin{widetext}
\bea
&&\int dx_1dx_2\Big[A_2\big\{g(x_1,x_2)\f{1}{\sqrt{2}}a^{\dg}_e(x_1)a^{\dg}_e(x_2)|0,1\ra+e(x_1)\delta(x_2)a^{\dg}_e(x_1)|0,2\ra+f(x_1)\delta(x_2)a^{\dg}_e(x_1)|0,3\ra\big\}+B_2\big\{j(x_1;x_2)\nn \\&&a^{\dg}_e(x_1)a^{\dg}_o(x_2)|0,1\ra+v(x_1)\delta(x_2)a^{\dg}_o(x_1)|0,2\ra+w(x_1)\delta(x_2)a^{\dg}_o(x_1)|0,3\ra\big\}+C_2~h(x_1,x_2)\f{1}{\sqrt{2}}a^{\dg}_o(x_1)a^{\dg}_o(x_2)|0,1\ra\Big] 
\label{wavefn}
\eea 
\end{widetext}
with $g(x_1,x_2)=g(x_2,x_1)$ and $h(x_1,x_2)=h(x_2,x_1)$. $e(x),f(x)~(v(x),w(x))$ are the probability amplitudes of one photon in the $e~(o)$ subspace while the 3LE in the excited state is $|2\ra$ and $|3\ra$, respectively. Here $A_2,B_2$, and $C_2$ identify the boundary conditions for the incoming photons. When both the photons are incoming from the left, $A_2=B_2=C_2=1/2$. Note that we express the two-photon scattering eigenstate in the space of free photons as well as the 3LE.  This is required to calculate a two-photon current in the system \cite{Roy10}. We evaluate various (seven) amplitudes in Eq.(\ref{wavefn}) by solving seven linear coupled first-order differential equations, which are obtained  from the stationary two-photon Schr{\"o}dinger equation. 
Some parts of $e(x)$, $f(x)$ (for $x>0$) and $g(x_1,x_2)$ (for $x_1,x_2>0$) fall exponentially with increasing $|x|$ or $|\tilde{x}|$ (with $\tilde{x}=x_1-x_2$), these are contributions from the two-photon bound state arising in the $ee$-subspace. We find in the $ee$ subspace for $x_c=(x_1+x_2)/2$,
\begin{widetext}
\bea
&&g(x_1,x_2)=\f{1}{2\pi\sqrt{2}}(\varphi_{k_1}(x_1)\varphi_{k_2}(x_2)+\varphi_{k_2}(x_1)\varphi_{k_1}(x_2))-\f{iVe^{i E_{\bk} x_c}}{\sqrt{2}}\Upsilon_1 e^{i(E_{\bk}-2E_2)|x|/2}e^{-(\gamma_2+\Gamma)|x|/2}\theta(x_1)\theta(x_2)\nn\\
&&e(x)=\f{1}{2\pi i V}\Big( \varphi_{k_1}(x)(1-\tau_{k_2})+\varphi_{k_2}(x)(1-\tau_{k_1})\Big) +\Upsilon_1  e^{i\Xi~ x}\theta(x)\nn\\
&&f(x)=\f{1}{2\pi}\big(\varphi_{k_1}(x)\rho_{k_2}\tilde{\rho}_{k_2}+\varphi_{k_2}(x)\rho_{k_1}\tilde{\rho}_{k_1}\big)+\Upsilon_2e^{i\Xi~ x}\theta(x)~~~~~{\rm where}\label{ampl}
\eea
\bea
\rho_k&=&\f{V}{k-E_2+i(\gamma_2+\Gamma)/2},~~\tilde{\rho}_k=\f{\Omega/2}{k-E_2+\Delta+i\gamma_3/2}, ~~\varphi_k(x)=e^{ikx}(\theta(-x)+\tau_k\theta(x))\nn \\\tau_k&=&1-iV\rho_k\big(1+\f{\Omega}{2V}\rho_k \tilde{\rho}_k\big),~~\Xi=(E_{\bk}-E_2)+i(\gamma_2+\Gamma)/2,~~\Upsilon_1=\f{1}{i\pi V}(1-\tau_{k_1})(1-\tau_{k_2})\nn \\
\Upsilon_2&=&\big(\rho_{k_1}\tilde{\rho}_{k_1}(1-\tau_{k_2})+\rho_{k_2}\tilde{\rho}_{k_2}(1-\tau_{k_1})\big)/2\pi\nn
\eea
\end{widetext}
We keep terms upto $\Omega^2$ order in the scattering state. Similarly we derive $j(x_1;x_2),~v(x),~w(x)$ in the $eo$ subspace. As there is no scattering in the $oo$-subspace, $h(x_1,x_2)=\phi_{\bf k}(x_1,x_2)/(2\sqrt{2}\pi)$. In our above calculations we start with a small finite $\gamma_3$, and the limit $\gamma_3 \to 0$ can be taken later. The amplitudes of the two-photon scattering state in Eq.\ref{ampl} in the limit $\Omega, \gamma_2, \gamma_3 \to 0$ match with the results for a 2LE coupled to a 1D waveguide \cite{Roy10}. 
The two-photon scattering state of the driven 3LE shows higher photon-photon correlation than that of the 2LE due to the presence of the classical driving field $\Omega$. It will be shown later explicitly from a two-photon current in these systems. The two-photon bound state in the scattering state also shows a strong  correlation between photons
; this is due to the 1D feature of scattering, and is similar for a 3LE and a 2LE. 

 Now we integrate out the field operators of the 3LE from Eq.\ref{wavefn}, and  write down an asymptotic outgoing scattering state in the original RR, LL and RL subspaces of free photons as a combination of two transmitted, two reflected and one transmitted plus one reflected photon.
\bea
&&\int dx_1dx_2\Big[t_2(x_1,x_2)\f{1}{\sqrt{2}}a^{\dg}_R(x_1) a^{\dg}_R(x_2)+r_2(x_1,x_2)\nn \\&&\f{1}{\sqrt{2}}a^{\dg}_L(x_1) a^{\dg}_L(x_2)+rt(x_1,x_2)a^{\dg}_R(x_1) a^{\dg}_L(x_2)\Big]|0,1\ra~,\label{asymstate} 
\eea
where $t_2,~r_2$, and $rt$ can be expressed in terms of $g(x_1,x_2),~h(x_1,x_2)$ and $\tau_k$.
A possible experimental set-up to measure $t_2$, $r_2$ and $rt$ has been proposed in \cite{Shenb07} by placing a beam splitter at the ends of the 1D waveguide with a single-photon counter  on each output arm of the beam splitter. 
For $\Omega=0$, the level $|3\ra$ turns  inactive, and it corresponds to a 2LE coupled to a 1D waveguide. 

A two-photon resonance in  the 2LE waveguide occurs for energy of the two incoming photons, $E_{k_1}=E_{k_2}$ and $E_{\bf k}=2E_2$. $|t_2(x_1,x_2)|^2$ has a maximum at $\tilde{x}=0$ at the resonance, and the magnitude of the maximum reduces with increasing $|E_{k_1}-E_{k_2}|$. $|t_2(x_1,x_2)|^2$ decays rapidly to zero with increasing value of $\tilde{x}$. The peak manifests bunching of two transmitted photons after scattering by an emitter. We find that the two-photon correlations in  a 2LE waveguide \cite{Shenb07} and a driven 3LE waveguide are quite similar at energies away from the TPR condition. 

\begin{figure}[t]
\includegraphics[width=8cm]{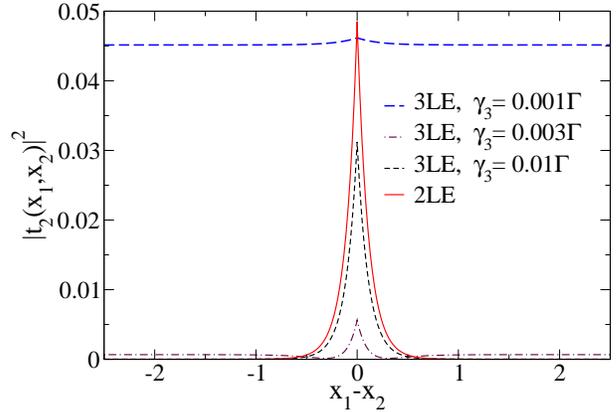}
\caption{ $|t_2(x_1,x_2)|^2$  of a driven 3LE $(\Omega=0.033\Gamma)$ and a 2LE $(\Omega=0)$ for  $E_{k_1}=E_{k_2}=E_2-\Delta$ and $\Delta=0.056\Gamma, \gamma_2=0.01\Gamma, \Gamma/2\pi=1.43$ MHz, and $E_{k},E_2$ are order of GHz. $x_1,x_2$ are scaled by $\Gamma$.}\label{pl2}
\end{figure}

In Fig.\ref{pl2} we plot $|t_2(x_1,x_2)|^2$ as a function of $\tilde{x}~(\equiv x_1-x_2)$ for energy of the incident photons, $E_{k_1}=E_{k_2}=E_2-\Delta$, at the TPR where the single-photon transmission and reflection coefficients of the 3LE waveguide show the EIT line shapes. $|t_2(x_1,x_2)|^2$ remains almost constant with increasing $\tilde{x}$, and has a large magnitude for the driven 3LE. It shows that the two transmitted photons are always bunched together at any distance separation. This may have significant practical application in quantum information processing to create far separated entangled photon pairs. $|t_2(x_1,x_2)|^2$ starts to decay with increasing $\tilde{x}$ for a deviation in energy of any of the two incident photons from the resonance condition. $|t_2(x_1,x_2)|^2$ for a 2LE waveguide has completely different behavior at the TPR as shown in the Fig.\ref{pl2}. We find that the line-shape of $|t_2(x_1,x_2)|^2$ of a driven 3LE waveguide starts to match with that of a 2LE waveguide as $\gamma_3$ is increased from 0 to a finite value (see Fig.\ref{pl2}). This is consistent with the fact that the single-photon EIT line-shapes of the driven 3LE waveguide are completely washed away for increasing $\gamma_3$. 

\begin{figure}[t]
\includegraphics[width=8cm]{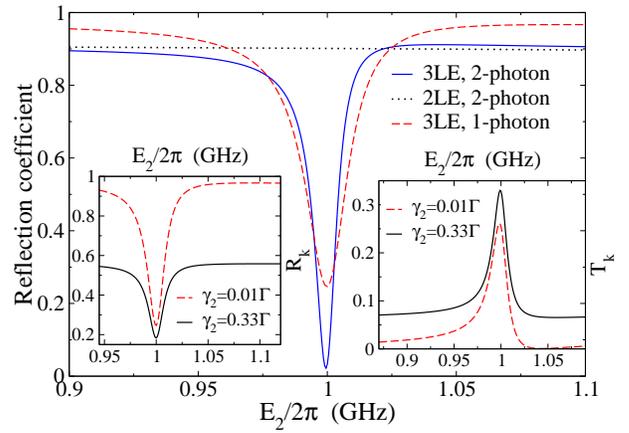}
\caption{Renormalized two-photon and single-photon reflection coefficients of a driven 3LE and a 2LE with $E_2$ for $\Delta=0.056\Gamma,\Omega=0.033\Gamma, \gamma_2=0.01\Gamma, \gamma_3=0.001\Gamma, \Gamma/2\pi=1.43$ MHz, and $E_k+\Delta=2\pi$ GHz. The insets show the single-photon reflection and transmission of a driven 3LE for different $\gamma_2$, and the other parameters are same as the main figure.}
\label{pl3}
\end{figure}

For a side-coupled emitter model, an expectation of the difference of photon number between right-moving and left-moving photons gives an estimate of total two-photon reflection coefficient (or two-photon reflection current). Total two-photon reflection current includes contributions of two reflected as well as one reflected plus one transmitted photons. Physically this is much easier to measure than $t_2,~r_2$, and $rt$;  one just needs to put a photo detector at the backward direction of the waveguide. We define the two-photon reflection current as
\bea
I&=&-\f{1}{2}\f{d}{dt}(N_R-N_L)~~{\rm where}~N_{i=R,L}\equiv \int dx a^{\dg}_i(x)a_i(x),\nn \\
&=&-\f{i}{2}[\mathcal{H},N_R-N_L]=\f{iV}{2}(a^{\dg}_o(0)|1\ra\la2|-|2\ra\la1|a_o(0)).\label{currentop}
\eea
The expectation of $I$ in the full two-photon scattering state has two parts, one is order of $\mathcal{L}$ and the other is order of $\mathcal{L}^0$ \cite{Roy10, Dhar08}, where $\mathcal{L}$ is roughly the length of the finite 1D waveguide. We find 
\bea
\la I \ra&=&\Big[\f{\mathcal{L}}{32\pi^2}(1-\tau_{k_1})\big(\f{3}{2}+\f{|\tau_{k_2}|^2}{2}\big)\nn \\ &-&\f{\tau^*_{k_1}\Upsilon_1\rho_{k_2}}{16\pi}\Big]+(1 \leftrightarrow 2)+{\rm H.c.}\label{curr}
\eea
By taking the limit $\Omega \to 0$ in Eq.\ref{curr} we get the corresponding current for a 2LE coupled to 1D waveguide, and the term of order $\mathcal{L}$ for the 2LE waveguide is just the contribution from two noninteracting photons. But, we can not separate the term of order $\mathcal{L}$ in Eq.\ref{curr} for the driven 3LE as a sum of the single-photon reflection coefficients $R_{k_1}$ and $R_{k_2}$. Therefore, the two-photon reflection current shows 1 order of magnitude higher photon-photon correlation for the driven 3LE compared to the 2LE case. We obtain a renormalized two-photon reflection coefficient for an average single photon from $\la I \ra$ after multiplying it by $\pi$. We use $\mathcal{L}\simeq 2\pi$ for an infinite waveguide. We plot the renormalized two-photon reflection coefficient for $\Omega=0$ (2LE) and $\Omega \ne 0$ (3LE) in Fig.\ref{pl3}. We find that the two-photon correlation has lowered the otherwise smaller dip in $R_k$ at the TPR. Note that $R_k$ does not vanish at the TPR as we use $\gamma_3\ne 0$.  Here we could not calculate a two-photon transmission current directly. The two-photon scattering in 1D induces inelastic scattering \cite{Shena07,Shenb07, Roy10} which enhances the effective value of $\gamma_2$.  In the insets of Fig.\ref{pl3}, we show that the peak of $T_k$ and the dip in $R_k$ are respectively increased and lowered as $\gamma_2$ is increased. Thus, we expect by comparing the line shapes of the single-photon and the two-photon reflection coefficients that the two-photon transmission  would be higher compared to the single-photon transmission $T_k$ at the TPR. The effect of the two-photon scattering on the reflection and the transmission coefficients is a little bit different for a strong driving field at $\gamma_3=0$.

In conclusion, we have presented a detailed investigation on the two-photon correlation in the driven 3LE-waveguide using an open system approach based on the Bethe ansatz, and compared it with the 2LE-waveguide system. 
Perceiving recent progress in experiments with a single driven three-level atom in a cavity and free space, we hope that the phenomena discussed here will be experimentally observed in the near future. We plan to further extend  the present approach to different 3LE configurations as well as four level systems in 1D waveguide for studying entanglement of photons \cite{Togan10}.

The work has been funded by the DOE under Grant No. DE-FG02-05ER46204.

\end{document}